\documentstyle[11pt,newpasp,twoside,epsf]{article}
\def\edcomment#1{\iffalse\marginpar{\raggedright\sl#1\/}\else\relax\fi}
\reversemarginpar

\begin{document}
\title{High-Resolution Observations of PSR~B1828$-$11}
 \author{I. H. Stairs}
\affil{Department of Physics and Astronomy, University of British Columbia, 
6224 Agricultural Road, Vancouver, BC V6T 1Z1, Canada}
\author{D. Athanasiadis, M. Kramer, A. G. Lyne}
\affil{University of Manchester,Jodrell Bank Observatory, 
Macclesfield, Cheshire SK11 9DL, UK}

\begin{abstract}
We present high-time-resolution observations of the young precessing pulsar
B1828$-$11, which yield clues to the true beam shape and the fundamental
precession period.
\end{abstract}

\section{Introduction}

PSR B1828$-$11 is a young, 405-ms pulsar discovered in the 20-cm
pulsar survey at Jodrell Bank in the late 1980s (\cite{clj+92}).
Regular observations at Jodrell Bank over nearly 15 years have
revealed correlated changes in the pulsar's average pulse profile
shape and rotational parameters, with strong periodicities of roughly
1000, 500 and 250 days.  It appears likely that some form of
precession is setting the timescale for these changes (\cite{sls00}),
which continue steadily.

\section{Continued Evidence for Precession}

\begin{figure}[t]
\plotfiddle{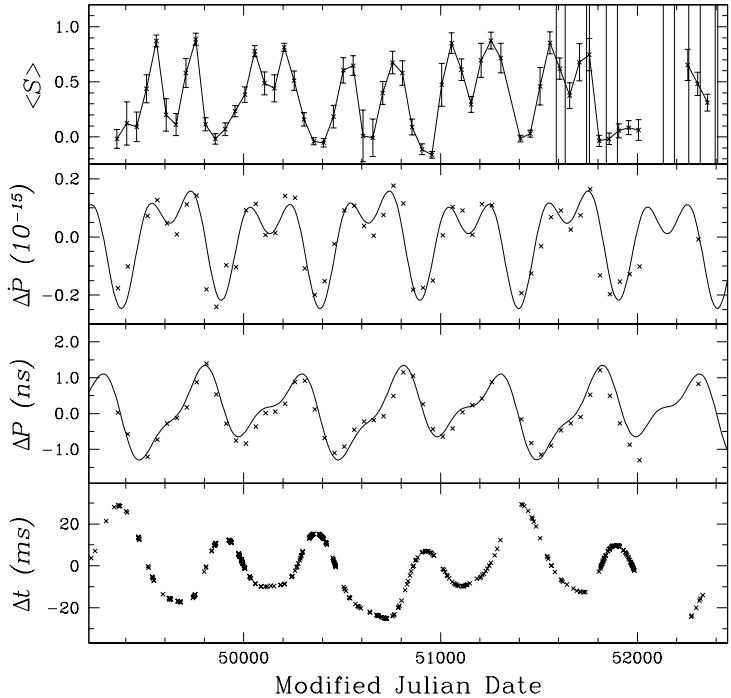}{4.2in}{0}{50}{50}{-150}{-80}
\caption{Residuals in arrival time $\Delta t$, period $\Delta P$ and
period derivative $\Delta \dot P$ relative to a simple spin-down
model, and the mean pulse shape parameter $S$ for over 3000 days of
observations.  The latter three series were calculated over time
intervals of 100 days which overlapped by 50 days.  The solid curves
show the predictions of a fit of three harmonically-related sinusoids
to $\Delta P$.  The vertical lines indicate the epochs at which Parkes
data were obtained.\label{fig:shapevar}}
\end{figure}

We describe the pulse profile from a given observation as a linear
combination of two templates, one slightly wider than the other.  The
resulting ``shape parameter'' is $S=A_{\rm N}/(A_{\rm N}+A_{\rm W})$,
where $A_{\rm N}$ and $A_{\rm W}$ are the fit heights of the narrow
and wide standard profiles respectively, so that $S\simeq1$ for 
narrow pulses and $S\simeq0$ for wider ones.  It is the average of
this parameter, along with the rotational quantities, which is
observed to vary with the phase of the precession
(Figure~\ref{fig:shapevar}).

In addition to the regular observations at Jodrell Bank Observatory
(gaps in the data in Figure~\ref{fig:shapevar} correspond to
maintenance or resurfacing of the 76-m telescope), we have been
acquiring high-resolution data with the 64-m Parkes telescope at
selected epochs throughout the current 1000-day precession cycle.
These data are taken with a $2\times512\times0.5$-MHz filterbank using
0.25\,ms sampling.  At the same time, we acquire correlator data with
which to study polarization.

\begin{figure}[t]
\plotfiddle{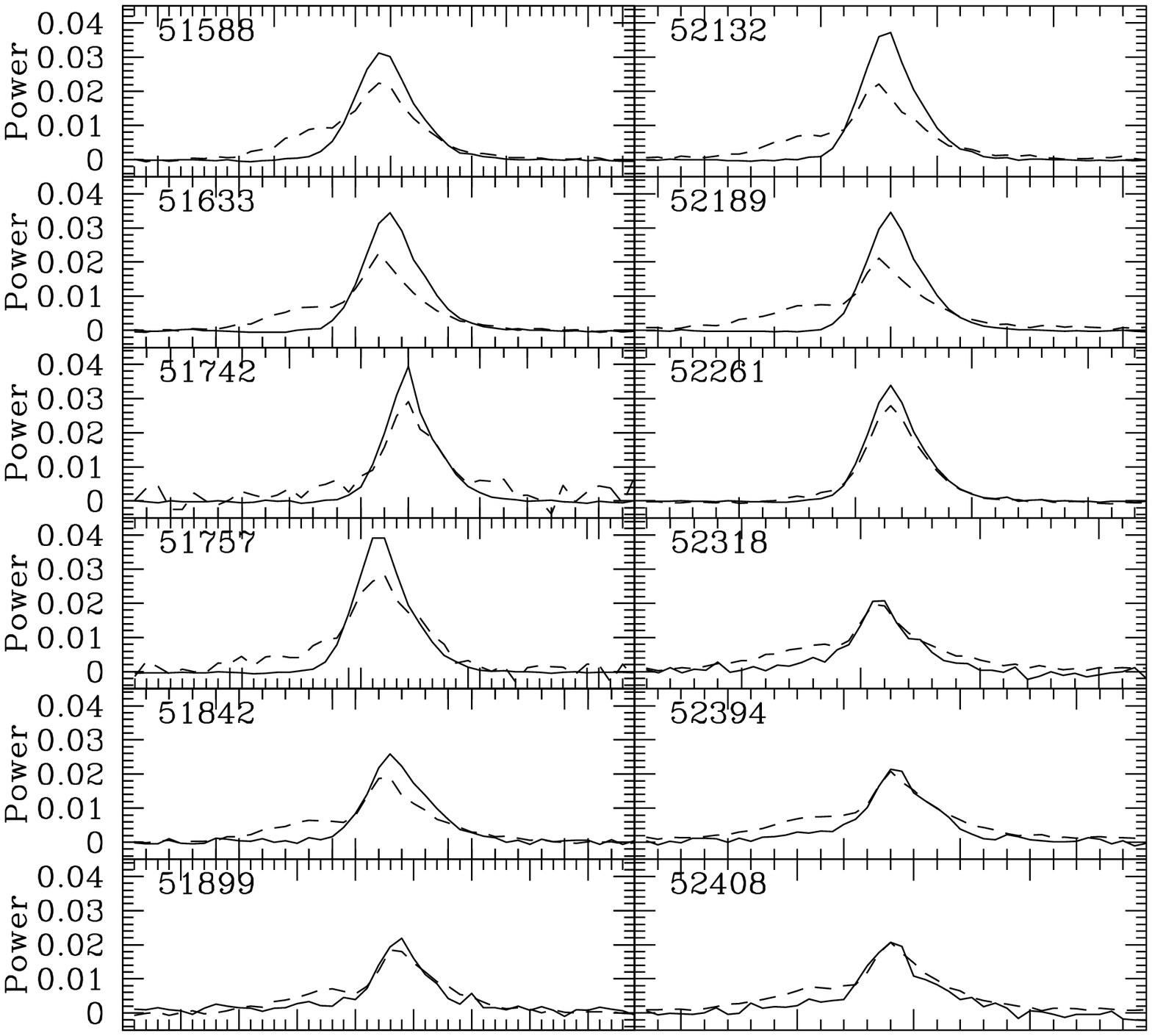}{3.3in}{0}{40}{40}{-130}{-80}
\caption{Narrow (solid) and wide (dashed) profiles at 12 different
Parkes observing epochs in 2000-2, produced by determining the shape
parameter $S$ for groups of 16 pulses.  The peak alignments are
arbitrary.  The flux normalization is arbitrary, though consistent
from epoch to epoch.\label{fig:widenarrow}}
\end{figure}

Unfortunately, individual pulses are too weak to be distinguished in
the Parkes data.  Instead, we add the time series into groups of 16
pulses, finding that this generally provides reasonable
signal-to-noise.  At each Parkes epoch, we separate these groups into
``wide'' and ``narrow'' sets using the shape parameter, and show the
averaged separated profiles in Figure~\ref{fig:widenarrow}.  These data
show that both narrow and wide modes are present at nearly all epochs,
indicating that mode-changing is occuring at each epoch, with some
evidence for changes in shape of the narrow and wide profiles.  There
is also good evidence for changes in average flux correlated with the
changes in average profile shape.  Overall the data appear to be
consistent with a combination of precession and mode-changing.  Future
work will determine whether the apparent differences in shape and
intensity at epochs separated by 500 days (for instance, MJDs 51757
and 52261) truly indicate a difference in pulse properties and hence a
fundamental precession period of 1000 days rather than 500 days, an
important issue in attempting to model the behaviour of this object
(e.g., \cite{le01,ja01,rez02}, Link, these proceedings).

\begin{figure}[t]
\plotfiddle{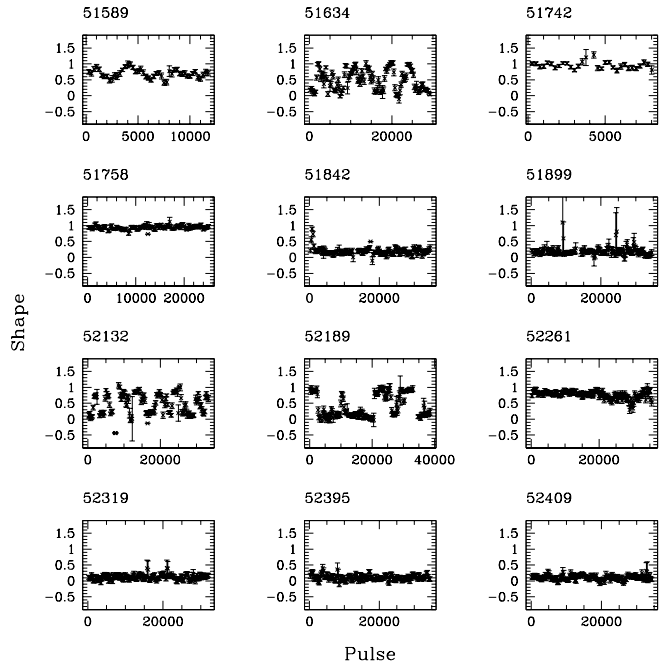}{3.5in}{0}{45}{45}{-140}{-80}
\caption{The shape parameter $S$ during each epoch of Parkes data,
averaged over 500 pulses, with bins overlapped by 250 pulses.  The
abrupt mode changes can be distinguished, and it is also evident that
the timescale for the mode changes varies dramatically with precession
phase.  Through Fourier analysis, significant periodicities can be
distinguished at the following MJDs: 51589 (3700 pulses), 51634 (7500
and 5000 pulses), 51742 (1100 pulses), 52132 (5000 pulses) and 52189
(7500 pulses). \label{fig:modes}}
\end{figure}

\section{Mode-Changing at Different Precession Phases}

In an attempt to study the mode-changing at each epoch more closely,
we calculate averages of the shape parameter $S$ in overlapping bins
of 500 pulses, using a method similar to that in
Figure~\ref{fig:shapevar}.  The results are displayed in
Figure~\ref{fig:modes}, where at the most ``active'' epochs
mode-switching can be seen to occur on timescales of a few thousand
periods, or tens of minutes.  In fact, significant periodicities can
be discerned at some of these epochs, as discussed in the caption to
Figure~\ref{fig:modes}.  We are constructing a 2-dimensional model of
the pulse beam shape and mode-changing properties, which will attempt
to address how these periodicities relate to the precession phase
and/or the average beam shape.

\section{Investigating Polarization Properties}

Another crucial clue to the 2-dimensional beam shape of this pulsar
lies in the polarization properties at different precession phases.
While careful calibration of the Jodrell Bank polarization data is
still in progress (see the contribution by Athanasiadis et al.), there
is already some evidence for a steeper position angle swing when wide
modes are strong.  Preliminary analysis of Parkes correlator
polarization data suggests similar behaviour, though instrumental
effects due to the linear telescope feeds need to be carefully
calibrated.  We will soon begin similar monitoring with the Green Bank
Telescope in order to achieve better signal-to-noise for small groups
of pulses, and to minimize instrumental effects in the observed
polarization.

\end{document}